# Consequences of niobium doping for the ferromagnetism and microstructure of anatase Co: TiO$_2$ films

S. X. Zhang[1], S. B. Ogale[1,*], L. F. Fu[2], S. Dhar[1], D. C. Kundaliya[1], W. Ramadan[1], N. D. Browning[2], and T. Venkatesan[1]

[1]Center for Superconductivity Research, Department of Physics, University of Maryland, College Park, MD 20742-4111.

[2]Lawrence Berkeley National Laboratory, NCEM, 1 Cyclotron Road, MS 72-150, Berkeley, CA 94720

Abstract:

It is shown that dilute niobium doping has significant effect on the ferromagnetism and microstructure of dilutely cobalt-doped anatase TiO$_2$ films. Epitaxial films of anatase TiO$_2$ with 3% Co, without and with 1% niobium doping were grown by pulsed-laser deposition at 875 $^0$C at different oxygen pressures. For growth at 10$^{-5}$ Torr niobium doping suppresses the ferromagnetism, while it enhances the same in films grown at 10$^{-4}$ Torr. High-resolution Z-contrast Scanning Transmission Electron Microscopy and Electron Energy Loss Spectroscopy show uniform surface segregation of cobalt-rich Ti$_{1-x-y}$Co$_x$Nb$_y$O$_{2-\delta}$ phase, but without cobalt metal clusters.

*Email: ogale@squid.umd.edu

Diluted magnetic semiconductors (DMS) have been a subject of intense research for the past few years due to the projected applications of these materials to spintronics and the potentially new mechanisms of ferromagnetism (FM) they seem to support[1-6]. Amongst the various classes of materials being pursued for intrinsic DMS-FM, oxides have perhaps been the most controversial. After the initial report of FM in dilutely cobalt doped anatase $TiO_2$ films by Matsumoto et al.[7], this system has been intensely researched by many groups around the world[8-17], and many other new systems have also been discovered[18-24]. These collective efforts are only now beginning to yield a better understanding of materials issues and raising the possibility of realization of an intrinsic DMS character in some systems. Recently it was shown that the FM in the $Co:TiO_2$ system can be significantly and reversibly modulated by an electric field applied through a ferroelectric gate[25]. Other studies[26] have established the presence of spin polarization in the system suggesting the role of carriers in the occurrence of FM.

The microstructure of $Co:TiO_2$ films and the growth sensitivity of the corresponding FM state observed in different works[8-17] possibly relates to the controllability of oxygen in the films. This has also led to an improved recognition of the relevance and possible significance of different defect types such as a neutral oxygen vacancy, an F-center etc. and the related new mechanisms[27, 28]. Since oxygen pressure influences film growth, the defect states as well as the carrier density through shallow donor character of oxygen vacancy in anatase $TiO_2$ has been difficult to isolate and control the corresponding effects in films. In this work we examine another channel to control the carrier density independently, namely by dual doping of niobium. We observe interesting changes in ferromagnetism as well as microstructure in this system.

Thin films of anatase $Ti_{1-x-y}Co_xNb_yO_{2-\delta}$ (x=0.03, and y=0.0, 0.01) studied in this work were grown on $LaAlO_3$ (LAO) by pulsed excimer laser deposition (KrF, $\lambda$=248 nm) at 875°C at different oxygen partial pressures. The laser energy density was ~ 2 J/cm² and the repetition rate was 10 Hz. The films were characterized by x-ray diffraction (XRD), Rutherford backscattering (RBS) channeling, four probe resistivity and Superconducting Quantum Interference Device (SQUID) measurements. The Z-contrast scanning transmission electron microscopy (STEM) and electron energy loss spectroscopy (EELS) experiments were carried out using a FEI Tecnai F20 UT microscope operated at 200 kV with a spatial resolution of 0.14 nm in scanning electron transmission mode (0.4 nm for EELS) and an energy resolution of 0.5eV. High-quality cross section electron microscopy samples were prepared using the shadow technique, a sample preparation technique that combines small angle cleavage technique with focused ion beam (FIB) minimizing sample modification.

Fig. 1a shows a typical XRD pattern observed for the $Ti_{1-x-y}Co_xNb_yO_{2-\delta}$ (x=0.03 and y=0.01) films. Only (00l) family of anatase film peaks are seen, in addition to the (LAO) substrate peaks. No major secondary contributions can be noted even on the log scale, suggesting high phase purity. The rocking curve on the main film peak (shown as an inset 1) has full width at half maximum (FWHM) of $0.22^0$ which shows the high degree of orientation order of the film. Similar general characteristics were seen for all the films grown under the stated conditions, indicating their optimal nature. It should be noted that small structural relaxations reflecting changes in the levels of strains were noted depending on the growth pressure as well as on doping. A typical effect of concentration dependence for $10^{-5}$ Torr case is shown in inset 2, where the difference of the (004) plane $d_1$-$d_2$ is about 0.007 Å.

Fig. 1b shows the RBS (Random and channeled) spectra for the $Ti_{1-x-y}Co_xNb_yO_{2-\delta}$ (x=0.03 and y=0.01) film. The elemental positions are indicated on the spectra. The Ti minimum channeling yield ($\chi_{min}$%) of about 8% seen in this spectrum for a Nb doped film is possibly the smallest reported thus far in the Co: $TiO_2$ DMS system, suggesting very high dopant substitutionality and very small distortion disorder. Interestingly, the random and channeled RBS spectra for $Ti_{1-x-y}Co_xNb_yO_{2-\delta}$ (x=0.03 and y=0.0) film without niobium doping (not shown) indicate a higher distortion disorder with the Ti minimum channeling yield ($\chi_{min}$%) of almost 38%. These data indicate the interesting role of Nb dopant in rendering a higher quality film growth and enhancing the degree of dopant substitutionality.

Fig. 2a shows a typical transmission electron micrograph for the $Ti_{0.96}Co_{0.03}Nb_{0.01}O_{2-\delta}$ sample. The micrograph (which is deliberately shown on a coarser length scale to bring out the presence or absence of any distributed clusters) is clearly very smooth and featureless, and no clusters of any kind are seen. The high resolution micrograph of the interface (not shown) ensured the epitaxy and high interfacial quality of the film growth consistent with the RBS channeling data. In Fig. 2b we show the high resolution micrograph of the surface region of the film (which is of interest for reasons discussed below) and it shows a highly crystalline character. The distribution of Co in the film was studied by probing the film via tens of EELS line scans across the sample starting with recording one signal on the $LaAlO_3$ substrate and subsequently probing the film in 5 nm steps in direction of the film surface. The range of energy for the line scan experiments was from 150 eV to 870 eV in order to include the signals of niobium, titanium, oxygen, cobalt and lanthanum. A typical series of line scans covering oxygen, cobalt and Lanthanum contributions is presented in Fig. 2c. The single scans are shifted in the y-direction for clarity. Proceeding from lower to higher counts in the graph (as indicated by the arrow in Fig. 2a), the first signal has been recorded on the LaAlO$_3$ substrate and includes the O K-edge at 532 eV and the La M$_5$-edge at 832 eV etc. The oxygen signal for the following spectra changes owing to the change in material by crossing the interface and being recorded on the Cobalt doped TiO$_2$ film. In addition, the La M$_4$, M$_5$-edge disappears while the Nb and Co signals are present consistently in the film. Remarkably, the cobalt signal was found only in the top ~ 15 nm surface layer of the sample. Also, the white lines for the Co L$_{2,3}$-edge have been observed at 795 eV and 780 eV respectively implying an oxidized state of Co and the creation of oxygen vacancies in the film. Notably the oxygen signal in these surface layers is also different than the signal in the deeper layers of the film with its strength also reduced, implying enhanced oxygen vacancy concentration. Substitution of Co$^{2+}$ for Ti$^{4+}$ is indeed expected to add oxyegn vacancies for local charge neutrality[28, 29]. This suggests the presence of a uniform surface layer phase wherein cobalt is highly enriched as compared to the mean intended concentration. We should like to point out that in the Co: TiO$_2$ samples without doped Niobium the cobalt distribution is seen to be fairly uniform at least at the concentrations being considered here[30]. In Fig. 2d we show the EELS line scans (again appropriately shifted for clarity) for the Nb and Ti contributions. It can be seen that the niobium signal is fairly uniform across the films thickness, except for the surface layer wherein the signal diminishes. It seems that niobium doping affects the charge balance of the local complex comprising of the substituted cobalt and the nearby oxygen vacancy, thereby driving cobalt out towards the surface. Notably this concurrently leads to a drop of Niobium concentration in the surface region. In this sense cobalt and niobium may not be entirely compatible ions for uniform dual doping in anatase $TiO_2$.

We note that surface enrichment of cobalt has been reported by Chambers et al.[9,31] in their films grown by oxygen plasma assisted MBE, in the form of tiny surface clusters presumably of the $CoTiO_3$ type. In our case the surface phase is in the form of a uniform surface layer. High-resolution Z-contrast imaging throughout the film confirms that no clusters of the dopant are present. Given the fact that we do not see any XRD signatures corresponding to $CoTiO_3$ in spite of the rather uniform surface enrichment may suggest that the structural coordination in the surface layer is still one of anatase $TiO_2$ type. Moreover, the surface layer is crystalline as shown by the high resolutionTEM micrograph of Fig. 2b.

Fig. 3a compares the room temperature magnetic hysteresis for $Ti_{1-x-y}Co_xNb_yO_{2-\delta}$ (x=0.03 and y=0.0) and $Ti_{1-x-y}Co_xNb_yO_{2-\delta}$ (x=0.03 and y=0.01) films, i.e. for 3% cobalt doped anatase $TiO_2$ films without and with 1% Nb doping grown at an oxygen pressure of ~$10^{-5}$ Torr (Note that substrate signal has been subtracted). Although both samples exhibit ferromagnetism, Nb doping is observed to suppress the magnetization considerably. Indeed the saturation moment ($M_S$) and coercive field ($H_C$) values drop from ~ 0.38 $\mu_B$/Co and ~300 Oe to about ~ 0.15 $\mu_B$/Co and ~80 Oe, respectively. In order to bring out the possible role of the carriers (electrons) known to be contributed by Nb to anatase $TiO_2$, we compared the resistivity for the 3% Co:$TiO_2$ films without and with 1% Nb doping. The corresponding data are shown in Fig. 3b. As expected, it is seen that the resistivity of the film with Nb doping is orders of magnitude lower than the film without Nb, and the Nb doped film is also metallic as against the semi-insulating nature of the film without Nb. This observation of the influence of chemically induced carrier manipulation on ferromagnetism is interesting and raises the possibility of the role carriers in the mechanism of ferromagnetism. This case can be compared and contrasted with that of the electric field induced manipulation of ferromagnetism reported recently using a field effect (FET) configuration[25]. The chemical manipulation is easier from the processing standpoint as compared to the FET device fabrication if it enables realization of the desired property goal. The same is however irreversible in contrast to the reversible field effect manipulation. Also, as seen in the present case, introduction of an extra dopant can potentially modify the local charge and chemical balance and lead to dopant redistribution effects.

In order to explore the connection between oxygen stoichiometry, resistivity and magnetization, further we compared these physical properties for the $Ti_{1-x-y}Co_xNb_yO_{2-\delta}$ (x=0.03 and y=0.0) and $Ti_{1-x-y}Co_xNb_yO_{2-\delta}$ (x=0.03 and y=0.01) films grown at a higher oxygen pressure of $10^{-4}$ Torr. The magnetization data (Fig. 3c) exhibit a very interesting reversal of signal strength as compared to the case of films grown at $10^{-5}$ Torr. The Co:TiO$_2$ film without Nb doping loses all the FM (as shown in the inset to Fig. 3c), while the FM does occur for the Co:TiO$_2$ film with Nb doping, albeit with a reduced moment. The $M_S$ value for Nb doped Co:TiO$_2$ film grown at $10^{-4}$ Torr is ~0.07 $\mu_B$/Co, smaller by a factor of two as compared to that for the film grown at $10^{-5}$ Torr, namely ~0.15 $\mu_B$/Co. The value of $H_C$ on the other hand is not seen to change significantly. The resistivity data for the films grown at $10^{-4}$ Torr are shown in Fig. 3d. Growth at this higher pressure imparts a highly insulating character to the Co:TiO$_2$ film without Nb doping, making it impossible to make a proper four probe resistivity measurement (therefore data are not shown). The room temperature resistivity for the Co:TiO$_2$ film with Nb doping is also seen to increase from 0.0028 Ohm-cm for growth at 10$^{-5}$ Torr to 0.014 Ohm-cm for growth at 10$^{-4}$ Torr; yet the metallicity is seen to have been preserved.

The fact that Nb doping leads to significantly lower resistivity in films grown at 10$^{-5}$ Torr and 10$^{-4}$ Torr, emphasizes the shallow donor nature of the Nb level[32]. The electrons thus contributed by Nb doping appear responsible for the changes to the magnetization. The chemical role of Nb in causing surface enrichment of cobalt should then be compounded with its role of electron source in controlling the degree of these effects.

In conclusion, niobium doping effects on the ferromagnetism and microstructure of dilutely cobalt doped anatase TiO$_2$ films are studied. It is shown that Nb induces formation of a uniform cobalt rich Ti$_{1-x-y}$Co$_x$Nb$_y$O$_{2-\delta}$ surface phase, without any cobalt metal clusters. The properties of the corresponding doped films are shown to depend fairly sensitively on the growth conditions.

The financial support for this work by DARPA Spin S program and NSF-MRSEC grants DMR-00-80008 and DMR-05-20471 is gratefully acknowledged. The microscopy work (N. D. Browning) was supported by NSF grant number DMR-0335364 and performed in the National Center for Electron Microscopy at Lawrence Berkeley National Laboratory supported by The Department of Energy under contract number DE-AC02-05CH11231.

**References:**

[1]. H. Ohno, *Science* **281**, 951(1998).

[2]. S. A. Wolf, D. D. Awschalom, R. A. Buhrman, J. M. Daughton, S. von Molnár, M. L. Roukes, A. Y. Chtchelkanova, and D. M. Treger, Science **294**, 1488 (2001).

[3]. A. H. Macdonald, P. Schiffer and N. Samarth, Nature Materials **4**, 195 (2005).

[4]. H. Ohno, H. Munekata, T. Penney, S. von Molnár, and L. L. Chang Phys. Rev. Lett. 68, 2664 (1992); H. Ohno, A. Shen, F. Matsukura, A. Oiwa, A. Endo, S. Katsumoto, and Y. Iye, Appl. Phys. Lett. **69**, 363 (1996).

[5]. T. Dietl, Nature Materials **2**, 646 (2003).

[6]. A.M. Nazmul, S. Sugahara, M. Tanaka, Appl. Phys. Lett. **80**, 3120 (2002).

[7]. Y. Matsumoto, Makoto Murakami, Tomoji Shono, Tetsuya Hasegawa, Tomoteru Fukumura, Masashi Kawasaki, Parhat Ahmet, Toyohiro Chikyow, Shin-ya Koshihara, and Hideomi Koinuma, Science **291,** 854 (2001).

[8]. S.A. Chambers, S. Thevuthasan, R. F. C. Farrow, R. F. Marks, J. U. Thiele, L. Folks, M. G. Samant, A. J. Kellock, N. Ruzycki, D. L. Ederer, and U. Diebold, Appl. Phys. Lett. 79, 3467 (2001),

[9]. S. A. Chambers, T. Droubay, C. M. Wang, A. S. Lea, R. F. C. Farrow, L. Folks, V. Deline, and S. Anders, Appl. Phys. Lett. **82**, 1257 (2003).

[10]. P. A. Stampe, R. J. Kennedy, X. Yan, and J. S. Parker, J. Appl. Phys. **92**, 7114 (2002).

[11]. Y. L. Soo, G. Kioseoglou, S. Kim, and Y. H. Kao, P. Sujatha Devi, John Parise, R. J. Gambino, and P. I. Gouma, Appl. Phys. Lett. **81**, 655 (2002).

[12]. S. R. Shinde, S. B. Ogale, S. D. Sarma, J. R. Simpson, H. D. Drew, S. E. Loafland, C. Lanci, J. P. Biban, N. D. Browning, V. N. Kulkarni, J. Higgins, R. P. Sharma, R. L. Greene, and T. Venkatesan, Phys. Rev. B **67**, 115211 (2003); S. R. Shinde, S. B. Ogale, J. S. Higgins, H. Zheng, A. J. Millis, R. Ramesh, R. L. Greene, and T. Venkatesan, Phys. Rev. Lett. **92**, 166601 (2004).

[13]. D. H. Kim, J. S. Yang, K. W. Lee, S. D. Bu, T. W. Noh, S.-J. Oh, Y.-W. Kim, J.-S. Chung, H. Tanaka, H. Y. Lee, and T. Kawai, Appl. Phys. Lett. **81**, 2421 (2002).

[14]. J.-Y. Kim, J.-H. Park, B.-G. Park, H.-J. Noh, S.-J. Oh, J. S. Yang, D.-H. Kim, S. D. Bu, T.-W. Noh, H.-J. Lin, H.-H. Hsieh, and C. T. Chen, Phys. Rev. Lett. **90**, 017401 (2003).

[15]. N. H. Hong, J. Sakai, W. Prellier, and A. Hassini, Appl. Phys. Lett. **83**, 3129 (2003)

[16]. G.C. Han, Y.H. Wu, M. Taya, K.B. Lia, Z.B. Guoa, T.C. Chonga, J. Magn. Magn. Mater. **268**, 159 (2004)

[17]. A. Punnoose, M. S. Seehra, W. K. Park, and J. S. Moodera, J. Appl. Phys. **93**, 7867 (2003).

[18]. T. Fukumara, Z. Jin, M. Kawasaki, T. Shono, T. Hasegawa, S. Koshihara, and H. Koinuma, Appl. Phys. Lett. **78,** 958 (2001).

[19]. D. C. Kundaliya, S. B. Ogale, S. E. Lofland, S. Dhar, C. J. Metting, S. R. Shinde, Z. Ma, B. Varughese, K.V. Ramanujachary, L. Salamanca-Riba and T. Venkatesan, Nature Materials **3**, 709 (2004).

[20]. S. B. Ogale, R. J. Choudhary, J. P. Buban, S. E. Lofland, S. R. Shinde, S. N. Kale, V. N. Kulkarni, J. Higgins, C. Lanci, J. R. Simpson, N. D. Browning, S. Das Sarma, H. D. Drew, R. L. Greene, and T. Venkatesan, Phys. Rev. Lett. **91**, 077205 (2003).

[21]. Y. G. Zhao, S. R. Shinde, S. B. Ogale, J. Higgins, R. J. Choudhary, V. N. Kulkarni, R. L. Greene, T. Venkatesan, S. E. Lofland, C. Lanci, J. P. Buban, N. D. Browning, S. Das Sarma, and A. J. Millis, Appl. Phys. Lett. **83**, 2199 (2003).

[22]. S. N. Kale, S. B. Ogale, S. R. Shinde, M. Sahasrabuddhe, V. N. Kulkarni, R. L. Greene, and T. Venkatesan, Appl. Phys. Lett. **82**, 2100 (2003)

[23]. N. H. Hong, J. Sakai, W. Prellier, A. Hassini, A. Ruyter, and F. Gervais, Phys. Rev. B **70**, 195204 (2004).

[24]. N. H. Hong, J. Sakai and N. T. Huong and V. Brizé, Appl. Phys. Lett. **87**, 102505 (2005)

[25]. T. Zhao, S. R. Shinde, S. B. Ogale, H. Zheng, T. Venkatesan, R. Ramesh, and S. Das Sarma, Phys. Rev. Lett. **94**, 126601 (2005).

[26]. H. Toyosaki, T. Fukumura, K. Ueno, M. Nakano, M. Kawasaki, Jpn. J. Appl. Phys. 44, L896 (2005); G. Herranz, R. Ranchal, M. Bibes, H. Jaffres, E. Jacquet, J. L. Maurice, K. Bouzehouane, F. Wyczisk, E. Tafra, M. Basletic, A. Hamzic, C. Colliex, J.-P. Contour, A. Barthelemy, A. Fert, cond-mat/0508289

[27]. A. J. Kaminski and S. Das Sarma, Phys. Rev. Lett. **88**, 247202 (2002); Phys. Rev. B **68**, 235210 (2003); S. Das Sarma, E. H. Hwang, and A. J. Kaminski, Phys. Rev. B **67**, 155201 (2003).

[28]. J. M. D. Coey, A. P. Douvalis, C. B. Fitzgerald, and M. Venkatesan, Appl. Phys. Lett **84**, 1332 (2004); J. M. D. Coey, M. Venkatesan, and C. B. Fitzgerald, Nat. Mater. **4**,173 (2005).

[29]. S. A. Chambers, S. M. Heald, and T. Droubay, Phys. Rev. B **67**, 100401(R) (2003)

[30]. S. R. Shinde, S. B. Ogale, Abhijit S. Ogale, S. J. Welz, A. Lussier, Darshan C. Kundaliya, H. Zheng, S. Dhar, M.S.R. Rao, R. Ramesh, Y. U. Idzerda, N. D. Browning, T. Venkatesan (Unpublished)

[31]. S. A. Chambers, C. M. Wang, S. Thevuthasan, T. Droubay, D. E. McCready, A. S. Lea, V. Shutthanandan, and C. F. Windisch, Jr., Thin Solid Films **418**, 197 (2002). S.A.Chambers et al. Appl. Phys. Lett. **82**, 1257 (2003).

[32]. D. Morris, Y. Dou, J. Rebane, C. E. J. Mitchell, R. G. Egdell, D. S. L. Law, A. Vittadini, and M. Casarin, Phys. Rev. B **61**, 13445 (2000).

**Figure captions:**

Fig.1. (a) The θ-2θ XRD spectrum for a $Ti_{0.96}Co_{0.03}Nb_{0.01}O_{2-\delta}$ film on $LaAlO_3$ (001). Peaks labeled "S" correspond to the substrate. Inset 1 shows the XRD rocking curve for the film. Inset 2 shows the change in (004) peak position between Nb doped (solid curve) and without Nb doped (dashed curve) films (b) 1.5-Mev $He^+$ Rutherford backscattering (RBS) random, simulation and channeling spectra for the $Ti_{0.96}Co_{0.03}Nb_{0.01}O_{2-\delta}$ film.

Fig.2. (a) Cross sectional TEM images at large length scales of the $Ti_{0.96}Co_{0.03}Nb_{0.01}O_{2-\delta}$ film grown on $LaAlO_3$ (001) substrate. (b) High resolution TEM image for the Cobalt-rich surface layer of the film. (c) & (d) EELS line scans across the cross section of this sample.

Fig.3. (a) Magnetic hysteresis loops for the $Ti_{0.97}Co_{0.03}O_{2-\delta}$ (open circles) and $Ti_{0.96}Co_{0.03}Nb_{0.01}O_{2-\delta}$ (filled circles) films grown at oxygen pressures of $10^{-5}$ Torr; (b) Resistivity Vs temperature of the two samples in (a); (c) Magnetic hysteresis loop for the $Ti_{0.96}Co_{0.03}Nb_{0.01}O_{2-\delta}$ film grown at oxygen pressures of $10^{-4}$ Torr. The inset shows the nonmagnetic behavior (diamagnetism of substrate) of the $Ti_{0.97}Co_{0.03}O_{2-\delta}$ sample grown at $10^{-4}$ Torr; (d) Resistivity Vs temperature of the $Ti_{0.96}Co_{0.03}Nb_{0.01}O_{2-\delta}$ film grown at oxygen pressures of $10^{-4}$ Torr.

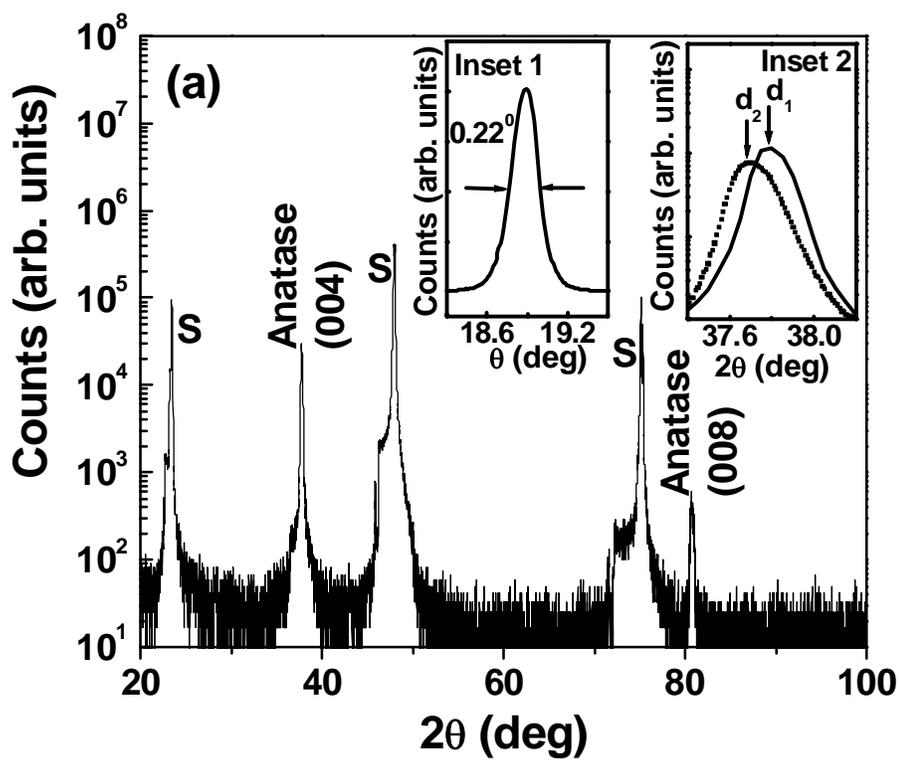

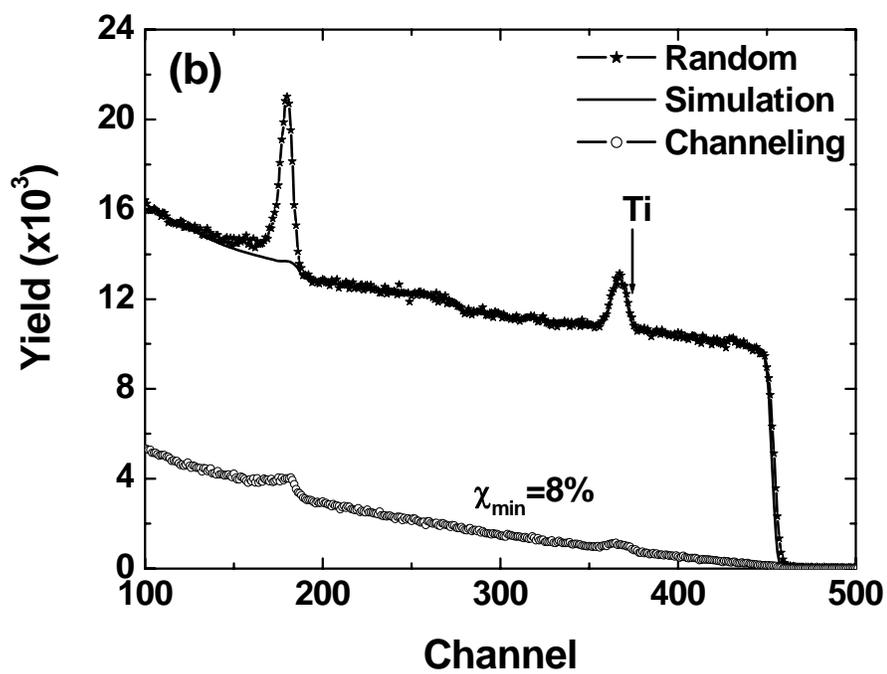

**Figure 1, Zhang et al.**

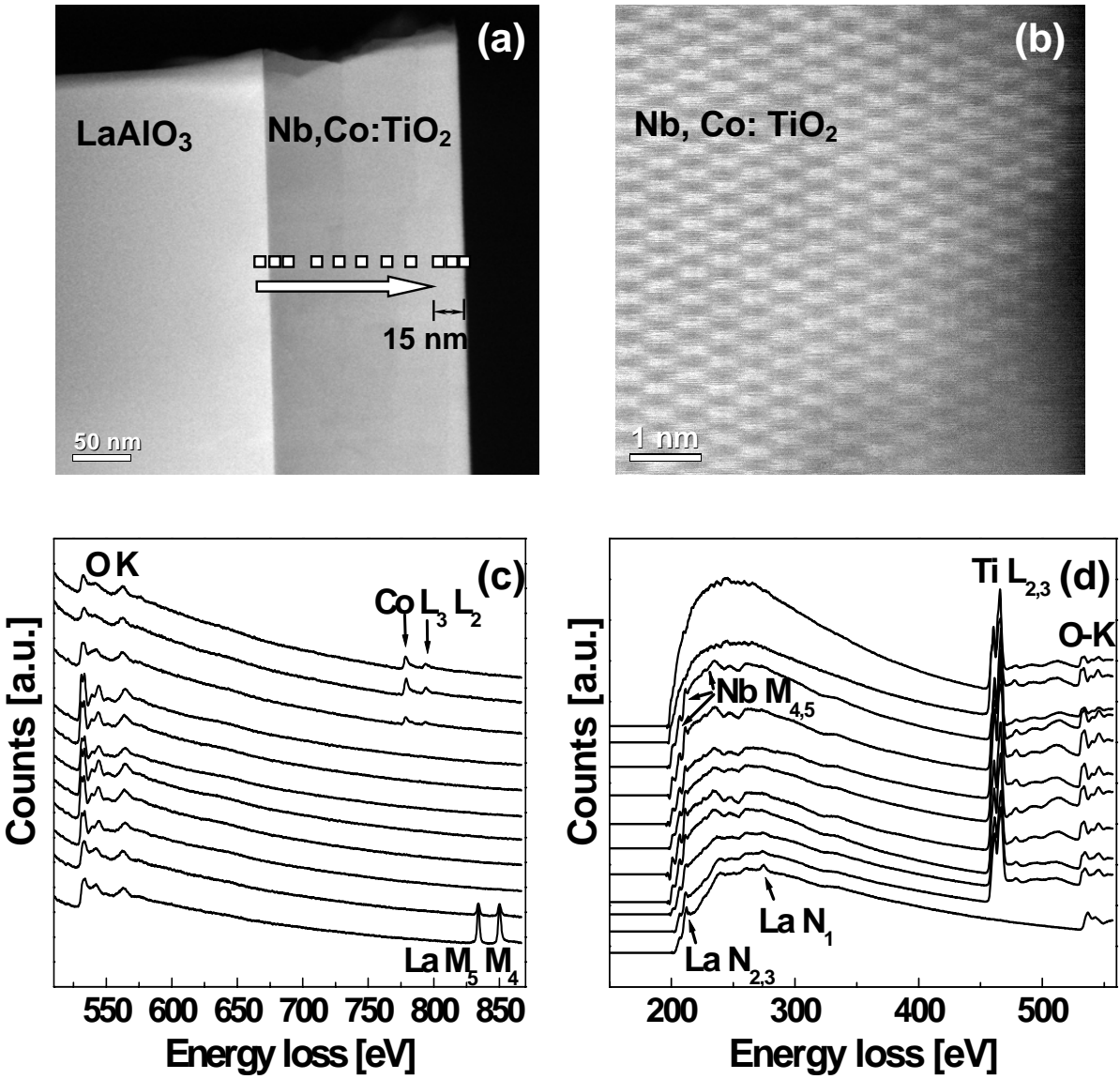

Figure 2 Zhang et al.

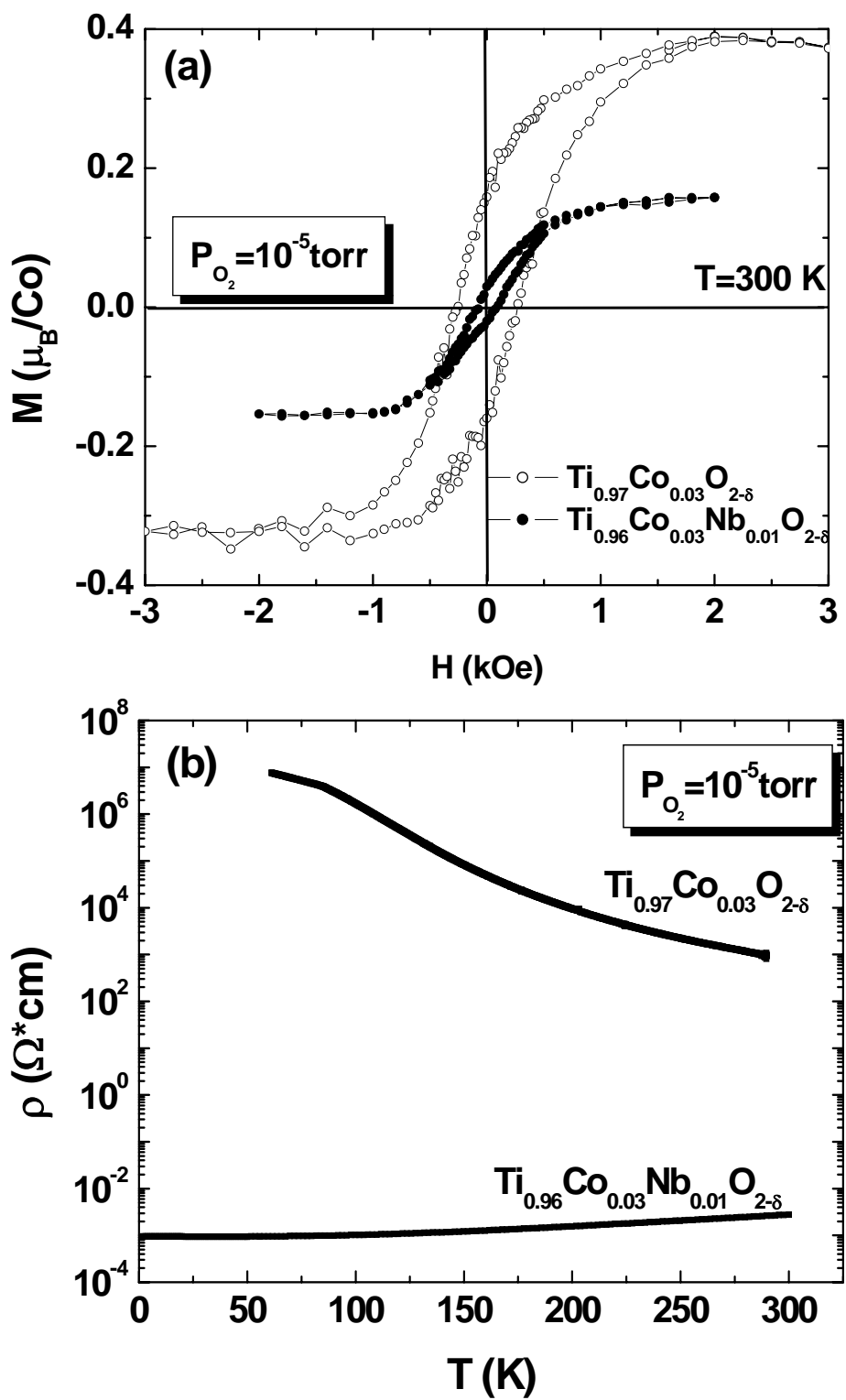

**Figure 3, Zhang et al.**

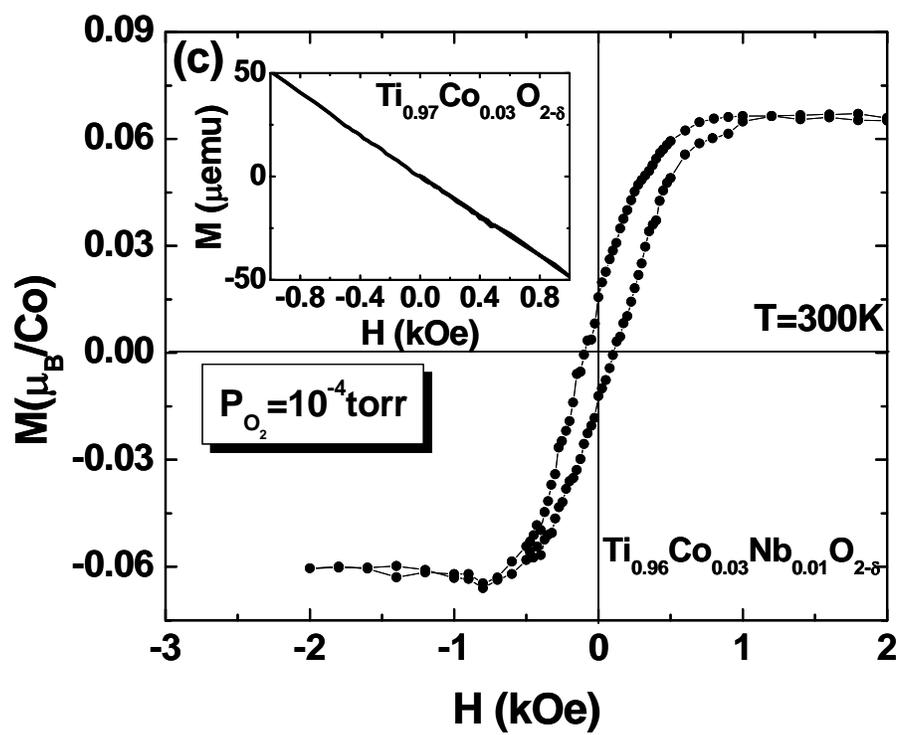

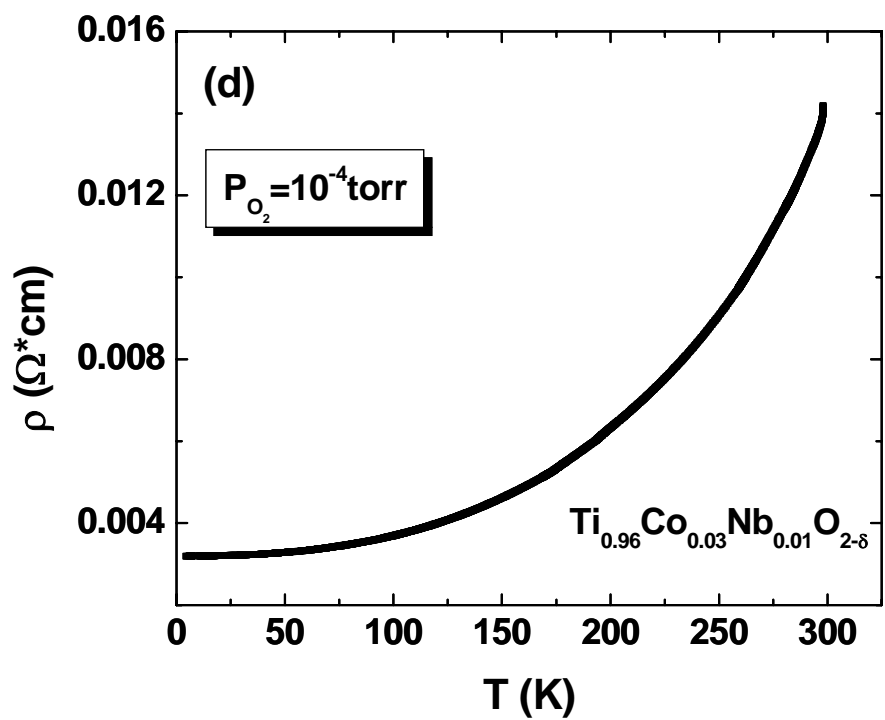

**Figure 3, Zhang et al.**